\documentclass[9pt,a4paper]{article}
\usepackage[a4paper,left=1.5cm,right=1.5cm,top=2cm,bottom=2cm]{geometry}
\usepackage{libertine}
\usepackage{authblk,multicol,cite,float}
\usepackage{amsmath,amssymb,amsfonts}
\usepackage[
   italic,
    symbolgreek
    ]{mathastext}
\usepackage{graphicx}
\usepackage{textcomp}
\usepackage[utf8]{inputenc}
\usepackage{indentfirst}

\newcommand{\subsubsubsection}[1]{\paragraph{#1}}

\usepackage{tikz}

\newcommand\copyrighttext{
  \footnotesize \textcopyright 2026 IEEE. Personal use of this material is permitted.
  Permission from IEEE must be obtained for all other uses, in any current or future
  media, including reprinting/republishing this material for advertising or promotional
  purposes, creating new collective works, for resale or redistribution to servers or
  lists, or reuse of any copyrighted component of this work in other works.}
\newcommand\copyrightnotice{
\begin{tikzpicture}[remember picture,overlay]
\node[anchor=north,yshift=-25pt] at (current page.north) 
  {\fbox{\parbox{\dimexpr\textwidth-\fboxsep-\fboxrule\relax}{\copyrighttext}}};
\end{tikzpicture}
}

\begin{document}
\title{Neurophysiological effects of museum modalities \\ on emotional engagement with real artworks}

\author[1]{Chen Feng}
\author[2,3]{Sébastien Lugan}
\author[3]{Karine Lasaracina}
\author[1]{Midori Sugaya}
\author[2]{Benoît Macq}
\affil[1]{\small Shibaura Institute of Technology, Tokyo, Japan}
\affil[2]{\small Université catholique de Louvain, Louvain-la-Neuve, Belgium}
\affil[3]{\small Royal Museums of Fine Arts of Belgium, Brussels, Belgium}

\date{}

\noindent\makebox[\textwidth][c]{
  \begin{minipage}{16cm}
    \maketitle
    \copyrightnotice
  \end{minipage}}
\noindent\makebox[\textwidth][c]{
  \begin{minipage}{15cm}
    \vspace{-5mm}
Museums increasingly rely on digital content to support
visitors’ understanding of artworks, yet little is known about how
these formats shape the emotional engagement that underlies meaningful
art experiences. This research presents an in-situ EEG study on how digital
interpretive content modulate engagement during art viewing.
Participants experienced three modalities: direct viewing of a Bruegel
painting, a 180° immersive interpretive projection, and a regular,
display-based interpretive video. Frontal EEG markers of motivational
orientation, internal involvement, perceptual drive, and arousal were
extracted using eyes-open baselines and Z-normalized contrasts.
Results show modality-specific engagement profiles: display-based
interpretive video induced high arousal and fast-band activity,
immersive projections promoted calm, presence-oriented absorption, and
original artworks reflected internally regulated engagement.
These findings, relying on lightweight EEG sensing in an
operational cultural environment, suggest that digital interpretive
content affects engagement style rather than quantity. This paves the
way for new multimodal sensing approaches and enables museums to
optimize the modalities and content of their interpretive media.
\footnote{This study has been made possible thanks to the ASEM-DUO Mobility
  Scholarship provided by ARES}
\end{minipage}}

~\\

\begin{multicols}{2}
\section{Introduction}
Art museums increasingly complement direct viewing of original works
with digital content such as immersive projections and
display-based videos
\cite{Trondle2012Physiology,Marin2019Real,Giorgi2023Virtual}.

While these formats differ in spatial scale and sensory framing, their
neurophysiological impact on emotional engagement remains poorly
understood. Emotional engagement is central to art experience yet not
directly observable, requiring validated neural markers such as
spectral power in delta, theta, beta, and gamma bands and frontal
alpha asymmetry
\cite{Kontson2015Your,Castiblanco2023Exploring,Vanutelli2025Predicting}.
Prior studies have focused mainly on behavioral outcomes
\cite{Herrera2017Modulation}, and comparisons between original
artworks and digital alternatives rarely include immersive projections
\cite{Marin2019Real}. Recent advances in wearable EEG
(electroencephalography) sensors enable systematic comparisons in
operational museum spaces \cite{Cruz2017Deployment}, but such studies
remain scarce.

The present study addresses this gap by examining three modes of art
experience (original artwork, immersive projection, and display-based
video) within a real museum environment.
By comparing engagement-related EEG markers across these modalities,
the study shows how different interpretive formats shape
emotional engagement with real artworks.

\section{Interpretive Strategies \\ in Art Viewing}

\subsection{Museum Context}

The Royal Museums of Fine Arts of Belgium launched the Bruegel Box in
2016 as part of the “Bruegel Unseen Masterpieces” project, offering
immersive projection experiences based on ultra-high-resolution
resources to present Bruegel’s works and related research in an
engaging format. It is the object of multiple scientific evaluations,
showing a sustained appreciation among the visitors, and became the focus
of an in-depth analysis as part of the SIAMESE\footnote{{\em Seamless
Image Archiving for Museums and Enhanced Sharing Experiences},
a FED-tWIN project supported by BELSPO}
research project. It provides a controlled, real-world setting
for physiological data collection, and was used in our study to deliver
a 180° panoramic interpretive projection alongside a display-based
video and direct viewing of the original painting. This tri-modal
configuration enables systematic comparison of how presentation
format (original artwork, planar video, or immersive
projection) modulates emotional and attentional responses during art
viewing.
This approach connects real-world museum practice with
neurophysiological evidence, offering a basis for understanding how
contemporary, digital interpretive content influences the visitor
experience. It introduces new insights into how these digital tools can
be optimized to enhance the museum experience and provide context for
artworks.

\subsection{Selection of the painting}

The painting {\em Winter Landscape With Skaters and a Bird Trap}
\footnote{Pieter Bruegel the Elder, {\em Winter Landscape with 
  Skaters and a Bird Trap}, Oil on oak panel, inv. 8724, Royal
Museums of Fine Arts of Belgium (Brussels)}
by Pieter Bruegel the Elder was selected as the central object
for this experiment due to its exceptional cultural and historical
relevance.
Bruegel’s winter scenes remain iconic in the cultural history of the
former Netherlands and are widely appreciated by the public, making
them an ideal choice for studies on audience engagement. 
The painting is documented in ultra-high-resolution formats,
enabling precise visual analysis and facilitating its integration into
digital experimental frameworks. It has been extensively studied
within the City and Society in the Low Countries project (Ghent
University, 2007--2017)\cite{meganck2018bruegel}.
To share these insights with the public, an immersive video was
created and is projected in the Bruegel Box,
aiming to deepen understanding of Bruegel’s artistic strategies and
socio-cultural context.
Beyond its apparent simplicity, the
composition reveals a semantic complexity that invites interpretive
engagement. While the scene initially conveys a sense of cheerfulness,
closer examination exposes latent elements of tension and foreboding,
reflecting Bruegel’s nuanced approach to themes of human existence and
social dynamics.

\section{EEG Indexes of \\ Emotional Engagement}

\subsection{Frontal Alpha Asymmetry (FAA)}

Frontal alpha asymmetry (FAA) is a long-established EEG marker of
relative frontal involvement, computed from the left-right balance of
alpha-band power, with alpha power inversely reflecting cortical
activation. Consistent evidence links left-dominant activity to
approach-oriented motivation and right-dominant activity to withdrawal
tendencies
\cite{Zhao2018Asymmetric,Chai2014Application,Kroupi2014EEG}.
FAA reliably varies with approach-withdrawal dynamics across
controlled studies of emotional images, reward anticipation, and
affective choice, and has been used in naturalistic contexts where
behavioural indicators of motivation are sparse
\cite{Olszewska2020Can}.

FAA is relevant to the present experiment for two methodological
reasons. (1) The three museum-relevant presentation
modalities (original artwork, immersive projection, and display-based
video) are expected to elicit distinct patterns of motivational
involvement even under matched visual content; FAA provides a
theory-grounded index for detecting such frontal-level
differences. (2) FAA can be stably obtained from lightweight,
frontal-only wearable EEG systems, making it feasible for in-gallery
field acquisition without disrupting visitor behaviour
\cite{Olszewska2020Can}.

Accordingly, FAA is incorporated as one of the study’s engagement-related EEG metrics to quantify modality-dependent differences in motivational allocation.

\subsection{Delta and Theta Activity}

Delta and theta activity are two of the most frequently examined low-frequency
components in EEG research. Although these bands are classically associated
with sleep-related processes in clinical contexts, studies in affective and
cognitive neuroscience have shown that delta and theta power also vary
systematically with states of internal processing, affective involvement, and
sustained task engagement \cite{Aftanas2002Time,Balconi2006EEG}. In particular, frontal theta increases have been
reported during tasks involving integrative evaluation, emotional processing,
and sustained attention \cite{Dmochowski2012Human}, while delta activity has been linked to conditions of
heightened internal demand or broad situational involvement \cite{Kontson2015Your}.

Because low-frequency power reflects large-scale fluctuations in processing
demand rather than precise oscillatory generators, delta and theta activity
can serve as coarse indexes of engagement in naturalistic environments \cite{Aftanas2004Analysis}.
Several studies using lightweight or wearable EEG systems have adopted
low-frequency activity as an index of general task-related involvement under
conditions where detailed behavioural markers are limited.

In the context of museum-based research, delta and theta activity are relevant
because they provide a means to capture broad differences in experiential
involvement across presentation modalities without requiring high-density
electrode coverage. Their interpretability in frontal wearable EEG makes them
suitable components of the engagement-related EEG indexes used in the present
study.

\subsection{Beta and Gamma Activity}

Beta and gamma activity are commonly examined high-frequency components in EEG
research and are generally associated with externally driven cognitive and
perceptual processes. In controlled laboratory settings, increases in beta and
gamma power have been observed during tasks requiring focused visual
inspection, perceptual discrimination, or sustained attentional control
\cite{Luther2023Oscillatory,Aydin2016Emotion}. These frequencies are also sensitive to properties such as stimulus contrast,
luminance, and visual complexity, making them responsive to bottom-up sensory
demands.

In naturalistic settings, high-frequency activity has been used as an index of
task-driven perceptual processing when precise behavioural control is not
feasible. Because beta and gamma power captures externally oriented attention
and perceptual load, these components offer a complementary perspective to the
more motivational or experiential indexes derived from FAA or low-frequency
activity.

For studies conducted in operational museum environments, beta and gamma
activity are relevant because presentation media can differ substantially in
illumination, field-of-view, and sensory framing. High-frequency activity thus
provides a set of engagement-related indexes that reflect the perceptual and
attentional demands imposed by different viewing conditions. However, gamma
activity can be weak or unreliable for non-foveal or naturalistic stimuli
\cite{Das2023Alpha}, and frontal beta suppression has been observed specifically during preferred
aesthetic experiences \cite{Herrera2017Modulation}.

\section{Methods}
\subsection{Participants}
Twenty healthy adult participants were recruited from the local
university and museum communities through announcements, bulletin
boards, and online advertisements. All participants were fluent in the
language of the study materials.
Participants were randomly assigned to one of two presentation groups
for the interpretive content conditions.

\subsection{Context of this study}
The study employed a mixed experimental design. All participants
viewed the original Bruegel painting in the gallery (original artwork
condition). Additionally, participants were randomly assigned to
experience one of two interpretive presentation modalities using the
same narrative content: (1) a screen-based introduction video (planar
display condition), or (2) an immersive projection introduction within
the Bruegel Box (immersive projection condition). This configuration
allows both within-subject comparisons (contrasting direct artwork
viewing with mediated interpretation) and between-subject comparisons
of how presentation modality (planar screen vs. immersive projection)
modulates emotional and attentional responses during art viewing.

\subsection{Experimental Procedure \\ and Condition Labeling}

Each viewing segment was labeled by modality (\emph{Original Artwork}, \emph{Immersive Projection}, \emph{Display-Based Interpretive Video}) and by block position (1, 2, or 3). Posture was coded as \emph{standing} for the Original Artwork condition and \emph{seated} for all interpretive content conditions. Presentation order was counterbalanced across participants. The primary statistical analyses focused on modality effects through within-subject comparisons (original artwork vs. interpretive content) and between-subject comparisons (immersive projection vs. planar display).

Ten participants were allocated to each of the two experimental
groups: immersive projection interpretive content paired with original
artwork viewing, and planar interpretive content paired with original
artwork viewing. This sample size was chosen to balance the
feasibility of recruitment and data collection within the museum
setting with the need for sufficient statistical power for both
within-subject and between-subject comparisons. While acknowledging
that this represents a minimal sample size for EEG and fNIRS
(functional near-infrared spectroscopy) studies,
it allows for a clear comparison of the primary experimental
conditions. We aimed for a balanced gender ratio within each group
where feasible, consistent with best practices in physiological
research. Future research may consider larger sample sizes to further
confirm the observed effects and enhance generalizability.

For all groups, neurophysiological responses (EEG and fNIRS) were 
continuously recorded throughout the viewing phases. After each viewing 
phase, participants completed a brief subjective questionnaire assessing 
their cognitive and emotional engagement, immersion (where applicable), 
and overall experience. The entire experimental session, including 
consent, setup, viewing, and debriefing, lasted approximately 60 to 
90 minutes. The experiment took place outside of the museum's opening
hours to ensure that the experience was perfectly reproducible for
each participant, who were free to withdraw at any time without penalty.

\subsection{Apparatus and Recording}
During our experiments, we recorded signals coming from four types of
sensors: EEG and fNIRS sensors, eye tracker\cite{Carter2020}
(tracking the direction of the gaze of the subject) and IMU
(accelerometers and gyroscopes tracking the movements of the head of
the subject).
However, this preliminary study focuses on the analysis
of the EEG signals.

EEG measures the electrical activity of the brain via electrodes
placed on the scalp.
It provides excellent temporal resolution, allowing for the precise
capture of rapid changes in brain states associated with cognitive
and emotional processing during the viewing phases. The EEG system
we selected for these experiments is the Muse S Athena, which
consists in a headband including four dry electrodes made of large
silver-thread fabric areas corresponding to locations TP9, AF7, AF8
and TP10. This procedure is comfortable and does not restrict
participant movement significantly during the viewing tasks.

All of the sensors used during our experiment are entirely
non-invasive and pose no known health risks to participants.

\begin{figure*}
  \center
  \includegraphics[width=\linewidth]{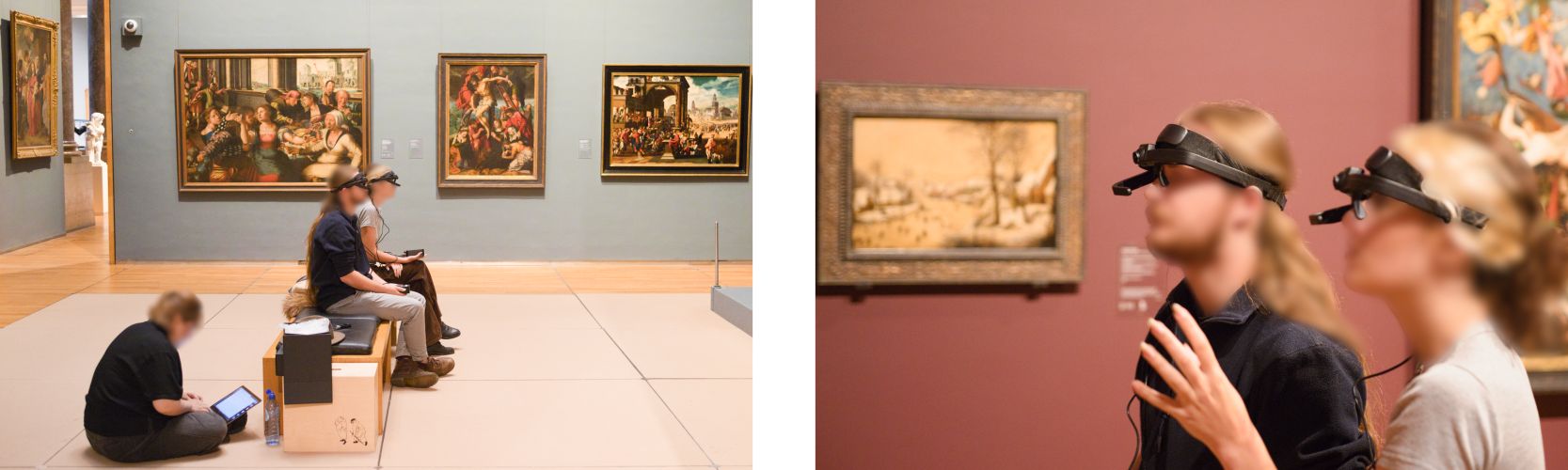}
  \caption{Experimental setting in the museums. Left: acquisition of
    the baseline with two subjects, right: experiment session in front
    of a real artwork}
\end{figure*}

The signals were streamed in real-time from the EEG/fNIRS sensors to
two smartphones (one associated with each participant) using BLE
(Bluetooth Low Energy). The smartphone were running MindMonitor\cite{9669778},
a mobile software to collect, pre-process, visualize and stream in
real-time data collected from the Muse EEG/fNIRS sensors.
The signals were then sent to a computer with real-time
recording software through a wifi router using the OSC (Open Sound
Control) UDP/IP protocol. The gaze direction within the field of view
of the participant were computed and recorded locally by the eye
trackers along with the video streams from the front facing camera as
well as the pupil camera. The gaze direction data were simultaneously
streamed in real-time using the MQTT TCP/IP protocol to the same
computer through the wifi router. All of the signals received from the
EEG, fNIRS, IMU and eye tracking sensors for both participants where
synchronized with a reference time-stamp.

A visual interface allowing the real-time visualization of the EEG and
fNIRS signals for both participants directly on the computer receiving
and recording the signals from all the sensors was built for this
experiment. This interface enabled the experimenters to ensure the
quality and validity (signals range, SNR, clear heartbeat visible on
the fNIRS signals, etc.) of the signals being recorded during each
activity.

The signals were simultaneously recorded and stored on the computer,
archived by sensor and participant with a common, synchronized
reference timestamp corresponding to the acquisition of the signals.

\subsection{Signal Processing \\ and Feature Extraction}

\subsubsection{Preprocessing} Analyses were performed on raw EEG
channels (TP9, AF7, AF8, TP10; 256Hz). Signals were band-pass filtered
(0.5--50Hz, zero-phase 5th-order Butterworth) with a 50~Hz notch
filter ($Q=30$) to suppress power-line interference.

Artifact detection combined device-level quality indexes and
signal-based criteria. Samples were excluded if
they met any of the following conditions: (i) flagged as low quality
by the device, (ii) exceeded accelerometer-based movement threshold
(95th percentile within segment), (iii) showed amplitude exceeding
$\pm 100~\mu\mathrm{V}$, or (iv) exhibited gradient changes exceeding
$50~\mu\mathrm{V}$/sample. Segments with fewer than 100 valid samples
were excluded. Independent component analysis (ICA) was applied to
remove identifiable ocular and motion artifacts when detected,
following conservative criteria for component selection.

\subsubsection{Spectral Decomposition} Band power was computed for
five frequency bands: delta (0.5--4Hz), theta (4--8Hz), alpha
(8--13Hz), beta (13--30Hz), and gamma (30--50~Hz). For each band and
channel, the signal was filtered to the target range and power was
calculated as mean squared amplitude
($\mu\mathrm{V}^{2}$). Channel-level values were averaged to obtain
segment-level band power estimates.
\subsubsection{Emotion-Related EEG Indexes}

\subsubsubsection{Frontal Alpha Asymmetry (FAA)}
FAA was computed from alpha power following Davidson’s approach.
Left-hemisphere alpha power was obtained by averaging TP9 and AF7;
right-hemisphere alpha power by averaging AF8 and TP10:
\begin{equation}
\mathrm{FAA} = \ln(\mathrm{Right}_{\alpha}) - \ln(\mathrm{Left}_{\alpha}).
\end{equation}
Positive values indicate relatively greater left-frontal activation
(approach tendency), whereas negative values reflect greater
right-frontal activation (avoidance tendency).

\subsubsubsection{Arousal Index}
Physiological arousal was quantified using the beta-to-alpha power ratio:
\begin{equation}
\mathrm{Arousal} = \frac{\mathrm{Beta}_{\mathrm{mean}}}{\mathrm{Alpha}_{\mathrm{mean}}}.
\end{equation}
Higher values indicate higher cortical arousal.

\subsubsection{Baseline Normalization}
Task-related changes were computed relative to the eyes-open (EO) baseline.
For FAA and the arousal index, the difference between the task value
and the EO baseline was used as the normalized metric.
Eyes-closed (EC) values were retained for descriptive comparisons.

\begin{figure*}
  \centering
  \includegraphics[width=\textwidth]{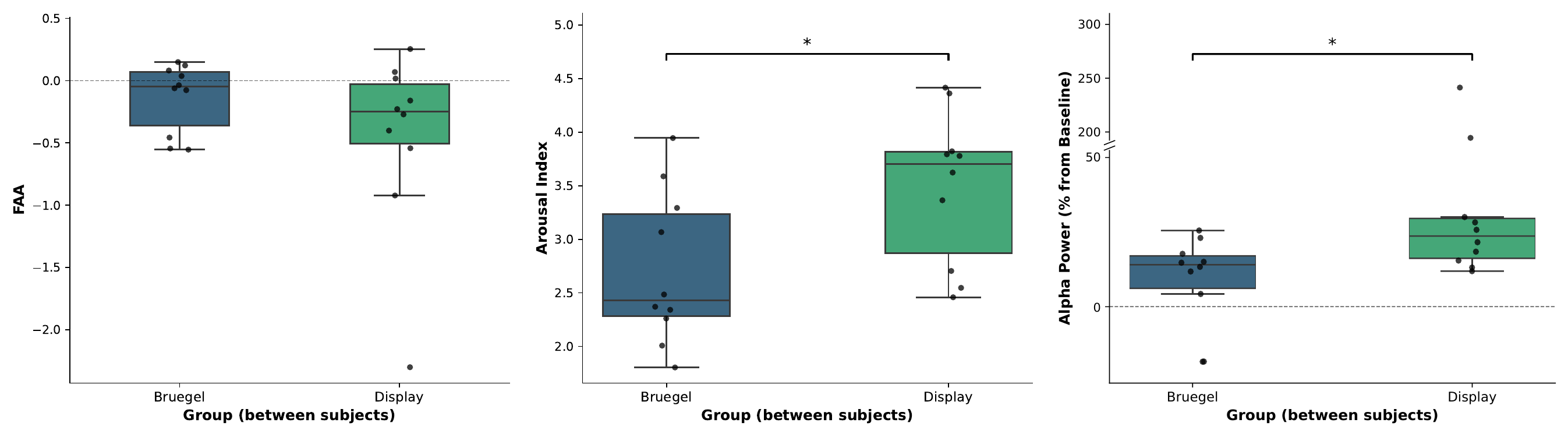}
  \caption{Left: frontal alpha asymmetry (FAA) across the immersive
    projection and display-based interpretive video; no reliable
    modulation by presentation modality was observed,
    center: between-subject comparison of the arousal index for the
    immersive wall-scale projection and the display-based interpretive
    video, showing higher arousal during display-based viewing,
    right: relative frontal alpha fraction for the immersive
    projection and display-based interpretive video, indicating a
    larger alpha proportion in the display condition.}
  \label{fig:group_1}
\end{figure*}

\begin{figure*}
  \centering
  \includegraphics[width=\textwidth]{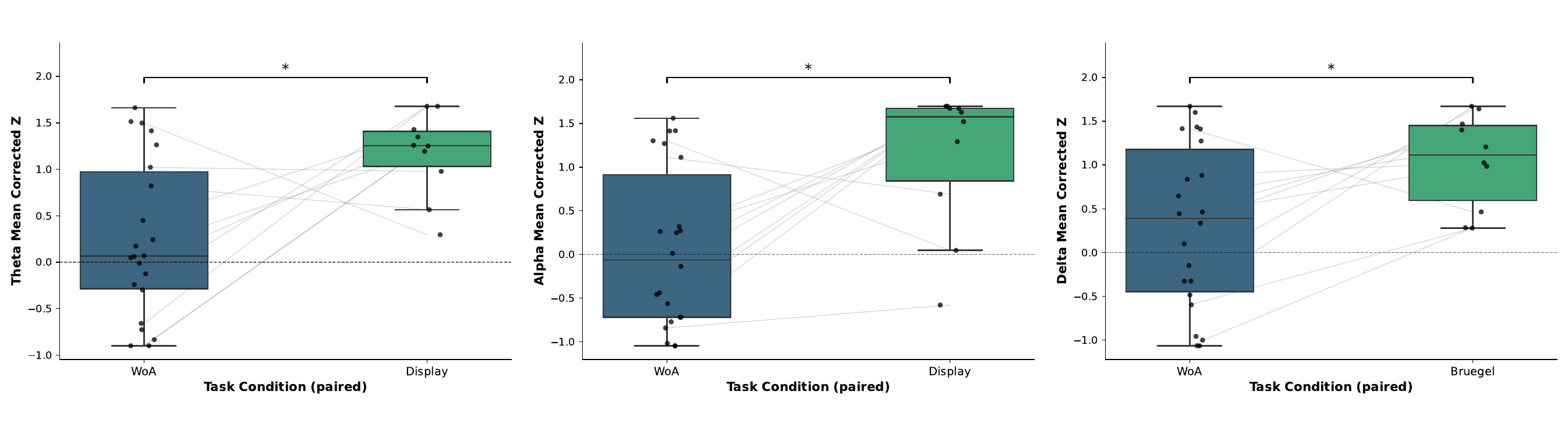}
  \caption{Left: within-subject comparison of Z-scored frontal theta
    power between original artwork (WoA) viewing and the display-based
    interpretive video,
    center: within-subject comparison of Z-scored frontal alpha power
    between original artwork (WoA) viewing and the display-based
    interpretive video,
    right: within-subject comparison of Z-scored frontal delta power
    between original artwork (WoA) viewing and the immersive
    wall-scale projection, showing selectively increased normalized
    delta in the immersive condition.}
  \label{fig:group_3}
\end{figure*}

\begin{figure*}
  \centering
  \includegraphics[width=\linewidth]{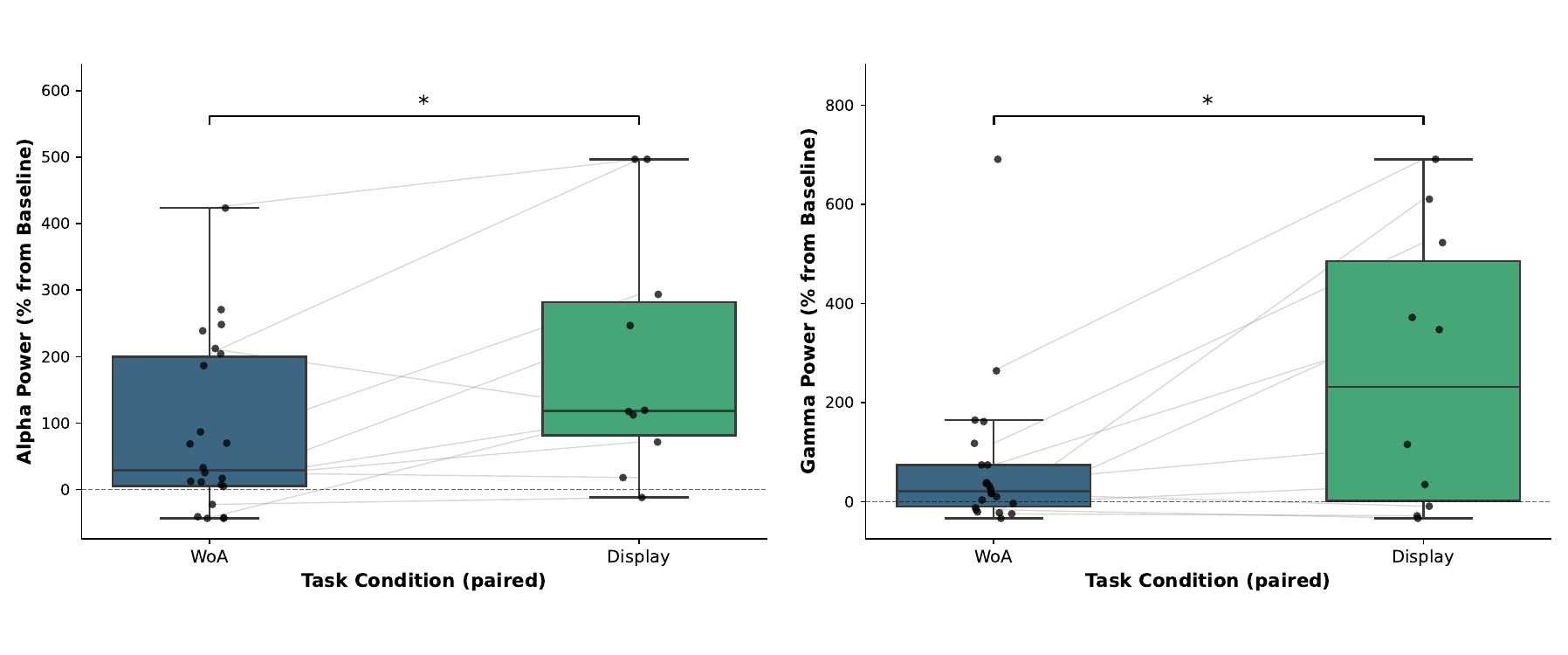}
  \caption{Left: within-subject comparison of baseline-corrected
    frontal alpha power between original artwork (WoA) viewing and the
    display-based interpretive video,
    Right: within-subject comparison of baseline-corrected frontal
    gamma power between original artwork (WoA) viewing and the
    display-based interpretive video.}
  \label{fig:group_2}
\end{figure*}

\section{Results}
This study examined how three museum-relevant presentation
modalities, the original artwork (WoA), an immersive wall-scale
projection, and a conventional display-based interpretive video,
shape emotional and cognitive engagement as reflected in frontal EEG.
Although frontal alpha asymmetry (FAA) did not differ by modality
(Fig.~\ref{fig:group_1}), the dataset revealed reliable task-dependent
differences in arousal and band-limited activity, indicating that
presentation formats bias the engagement style rather than uniformly
amplifying “emotion.”

\subsection{Externally Driven Arousal \\ During Display-Based Viewing}
A between-subject comparison showed a significantly higher
Arousal\_Index for the display-based interpretive video than for the
immersive projection (Display > Immersive, Welch’s $t$, $p=0.026$;
Fig.~\ref{fig:group_1}), whereas WoA occupied an intermediate range
without significant differences from either digital modality.
In within-subject contrasts against WoA, the display condition also 
showed baseline-corrected increases in alpha and gamma power
($p=0.043$ and $p=0.033$; Fig.~\ref{fig:group_2}) and higher Z-scored
theta and alpha ($p=0.045$ and $p=0.037$; Fig.~\ref{fig:group_2}), and
between subjects the display further exhibited a larger relative alpha
fraction than the immersive projection (Mann-Whitney, $p=0.021$;
Fig.~\ref{fig:group_1}), a constellation that aligns with widely
reported EEG signatures of stimulus-guided information uptake and
sustained attentional tracking during dynamic audiovisual sequences.
Because the visual content was matched across the two digital
modalities, these effects plausibly reflect differences in spatial
presentation and pacing rather than narrative content per se, and they
should not be over-interpreted as “deeper” aesthetic valuation. 

\subsection{Calming Immersive Presence \\ in Wall-Scale Projection}
The immersive projection did not show the heightened arousal observed
for the display, yet it differed from WoA in a normalized delta
measure (Delta\_Mean\_Corrected\_Z; immersive > WoA, paired test,
$p=0.025$; Fig.~\ref{fig:group_3}) while otherwise lacking broad
fast-band enhancement. Modest frontal delta increases in awake,
low-motion settings are commonly associated with quiet absorption and
reduced attentional competition, and the enclosing field-of-view with
diminished peripheral boundaries in immersive rooms plausibly supports
such a calmer, presence-oriented engagement.
Consistent with this, the immersive condition also showed a lower
relative alpha fraction than the display (between subjects, $p=0.021$;
Fig.~\ref{fig:group_1}), indicating that immersive presentation is not
simply a “stronger display,” but a qualitatively different mode of
orienting and settling into the imagery.

\subsection{Reflective, Self-Paced Engagement \\ in Original Artwork Viewing}
WoA did not exhibit the fast-frequency enhancements seen for the
display nor the normalized delta increase observed for the immersive
projection, and its Arousal\_Index did not differ significantly from
either digital modality.
This is best characterized as an absence of robust spectral power
deviations rather than “lowest arousal,” and is compatible with
accounts that original artworks often elicit reflective, internally
regulated, meaning-oriented processing that may not manifest as large
frontal power changes despite substantial subjective involvement.
Importantly, FAA showed no reliable modulation across modalities, so
approach-withdrawal tendencies should not be inferred to differ.

\subsection{Three Modality-Specific \\ Engagement Styles}
Synthesizing the arousal and spectral outcomes suggests three
engagement styles within the limits of frontal EEG: (i) the
display-based interpretive video engages a high-arousal, externally
driven information-uptake mode (Arousal\_Index higher than immersive;
multiple fast-band enhancements vs WoA; larger relative alpha than
immersive); (ii) the immersive projection supports a calmer,
presence-oriented mode marked by selectively increased normalized
delta vs WoA and no additional arousal elevation; (iii) original
artwork viewing promotes a reflective, self-paced mode without robust
spectral amplification, aligning with interpretive, internally
generative appraisal rather than externally paced sensory drive.
These modes are different affective-cognitive profiles, not a
hierarchy of “better” versus “worse” engagement.

\subsection{Methodological Considerations \\ and Future Directions}

Two analytic choices warrant emphasis to preempt misinterpretation:
first, we report both baseline-corrected powers and per-subject
Z-scores because absolute EEG power varies widely across individuals,
and Z-normalization mitigates inter-subject scaling differences to
better reveal relative task contrasts (the delta effect was
significant in Z but not in raw corrected power); second, the
display-immersive content was matched, so observed differences most
parsimoniously relate to spatial context and pacing rather than
narrative content.
More generally, frontal-only wearable EEG cannot resolve posterior
visual dynamics or deep affective evaluation, and within-subject
comparisons were limited to $N=10$ with between-subject contrasts at
$N=20$, which constrains sensitivity and generalizability.
Within the constraints of the current experimental design, the present
results nonetheless show that the original painting, immersive
projection, and display-based interpretive video differentially shape
arousal regulation and engagement style.
Future work should combine frontal and temporal fNIRS, eye-tracking,
and calibrated self-report to capture evaluative meaning-making and
immersive presence more comprehensively.

\section{Conclusion}
This study demonstrates that presentation modality in museum
contexts (original artworks, immersive interpretive projections, and
display-based videos) shapes distinct engagement profiles rather than
uniformly amplifying emotional involvement. EEG results indicate that
display-based videos elicit higher arousal and fast-band activity,
immersive projections foster calm, presence-oriented absorption, and
original artworks support reflective, internally regulated
engagement. These findings highlight that digital interpretive content
modulates the style of engagement, not its magnitude, and underscore
the feasibility of lightweight EEG sensing in operational cultural
environments. Future work should integrate multimodal measures
(EEG, fNIRS, eye tracking, self-report) to capture evaluative
meaning-making and immersive presence more comprehensively, advancing
pervasive computing applications for cultural heritage and experience
design.

\bibliography{refs}
\bibliographystyle{unsrt}
\end{multicols}

\end{document}